\documentstyle[preprint,aps,eqsecnum]{revtex}

\def\beq#1{\begin{equation} \label{#1}}
\def\eeq{\end{equation}}
\def\bra#1{\left\langle #1\right\vert}
\def\ket#1{\left\vert #1\right\rangle}

\input prepictex
\input pictex
\input postpictex
\newdimen\tdim
\tdim=\unitlength
\def\stpltsmbl{\setplotsymbol ({\small .})}

\newbox\phru
\setbox\phru=\hbox{\beginpicture
\setcoordinatesystem units <\tdim,\tdim>
\stpltsmbl
\setquadratic
\plot
0 0
2.5 3
5 0
7.5 -3
10 0
/
\endpicture}
\def\photonru #1 #2 *#3 /{\multiput {\copy\phru}  at
#1 #2 *#3 10 0 /}

\newbox\sru
\setbox\sru=\hbox{\beginpicture
\setcoordinatesystem units <\tdim,\tdim>
\stpltsmbl
\setquadratic
\plot
  0.0   0.0
  4.8   1.5
  7.5   5.0
  7.3   8.5
  5.0  10.0
  2.7   8.5
  2.5   5.0
  5.2   1.5
 10.0   0.0
/
\endpicture}
\def\springru #1 #2 *#3 /{\multiput {\copy\sru}  at
#1 #2 *#3 10 0 /}

\def\beq#1{\begin{equation} \label{#1}}
\def\eeq{\end{equation}}
\def\bra#1{\left\langle #1\right\vert}
\def\ket#1{\left\vert #1\right\rangle}

\begin{document}
{
\tighten
\begin{center}
{\Large\bf Entangled symmetries  explain without QCD dynamics\\
CP violation in $B^o \rightarrow K\pi$; not in $B^{\pm} \rightarrow K\pi$ decays
\\
Unexpected isospin relations in $B^{\pm}\rightarrow K\pi$ and $B^o\rightarrow K\pi$
}
\vrule height 2.5ex depth 0ex width 0pt
\vskip0.8cm
Harry J. Lipkin\,$^{b,c}$\footnote{e-mail: \tt ftlipkin@weizmann.ac.il} \\
\vskip0.8cm
{\it
$^b\;$School of Physics and Astronomy \\
Raymond and Beverly Sackler Faculty of Exact Sciences \\
Tel Aviv University, Tel Aviv, Israel\\
\vbox{\vskip 0.0truecm}
$^c\;$Department of Particle Physics \\
Weizmann Institute of Science, Rehovot 76100, Israel \\
and\\
High Energy Physics Division, Argonne National Laboratory \\
Argonne, IL 60439-4815, USA}

\end{center}

\vspace*{0.8cm}

\centerline{\bf Abstract}

Simple flavor symmetry argument without QCD dynamics shows why CP violation observed
in neutral $B$ to $K\pi$ decays is absent in charged $B$ decays where  tree diagram final state has two $u$ quarks satisfying Pauli principle. Entanglement preserves short range symmetry correlations after separation into two mesons. Pauli principle and symmetries require totally flavor-symmetric tree diagram final state. $\pi \pi$ isospin state with I=2 already is flavor symmetric and suffers no symmetry constraint.
Strange  flavor-symmetric state with V spin V=2 is linear combination of $K\pi$ and $K \eta$ with probability only (1/4) for $K\pi$. Tree diagram suppression by factor 4 not present in neutral decays explains  negligible tree-penguin interference and CP violation in charged decays. Detailed full symmetry analysis shows constraints from
space-inversion, charge conjugation, Pauli antisymmetrization and flavor symmetry.
Two antiquarks produced at same space point by tree diagram have even parity. Even parity final state requires even parity, even space symmetry and  color-spin antisymmetry for two $u$ quarks. Color-singlet spin-singlet final state requires color-spin antisymmetry and therefore flavor symmetry for two-antiquark wave function. Flavor symmetric $\bar u \bar d$  and $\bar u \bar s$ antiquark pairs have isospin I=1 and V-spin V =1.
Generalized charge conjugation  invariance requires I=2 for $\pi\pi$ tree diagram and V = 2 for $K\pi$.
Penguin decays give I = 1/2 final state and no CP violation. Experiments confirm surprising predictions from tree suppression factor not noted in previous analyzes.
I=1/2  violations seen only in relation between charged and neutral decays together with no I=3/2 components in each individual decay. Common treatments using  $\pi \pi$ data fail to fit $K\pi$ data.

\vspace*{4mm}

\vfill\eject
\section{Introduction - CP violation observed in neutral $B$ decays but not in charged decays}

Direct CP violation has been experimentally observed\cite{PDG,HFAG}
in $B_d \rightarrow K^+ \pi^-$ decays.
\beq{acp0}
A_{CP}(B_d \rightarrow K^+ \pi^-)= -0.098 \pm 0.013
\eeq
 The failure to observe CP violation in charged
decays\cite{Ali} has been considered a puzzle\cite{nurosgro,ROSGRO}
because changing the flavor of a spectator quark which does not participate
in the weak decay vertex is not expected to  make a difference.
\beq{acp+}
\begin{array}{ccl}
\displaystyle
A_{CP}(B^+ \rightarrow K^o_S \pi^+)= 0.009 \pm 0.029
\hfill\\
\\
A_{CP}(B^+ \rightarrow K^+ \pi^o)= 0.051 \pm 0.025
\end{array}
\end{equation}

A general theorem from CPT invariance shows\cite{lipCPT} that direct CP violation can occur only via the interference between two amplitudes which have
different weak phases and different strong phases. This  holds also for all
contributions from new physics beyond the standard model which conserve CPT.

$B\rightarrow K\pi$ decays are dominated by the penguin diagram shown in fig. l
with smaller contributions from the tree diagram shown in fig. 2.
CP violation is believed to arise from interference between the dominant penguin and smaller tree amplitudes.
The assumption that the tree amplitude is sufficiently small to be
treated in first order in the square of the total amplitude
has led to the ``Lipkin Sum Rule"\cite{Ali}.
The agreement with experiment\cite{PDG,HFAG} confirmed this assumption  \cite{approxlip,approxgr,Gronau,ketaprimfix}. However, these results gave no information about tree-penguin interference from data available at that time. The experimental errors were then so large that they could be fit by a pure penguin contribution

\section {Model-independent search for tree-penguin interference}

\subsection{Isospin analysis}

The search begins with a model-independent isospin analysis. The sum rule is most conveniently written \cite{ketaprimfix}
as a ``difference rule"
\beq{eqapp}
{{\tau^o}\over{\tau^+}}\cdot \left[ 2B(B^+ \rightarrow K^+ \pi^o)
- B(B^+ \rightarrow K^o \pi^+ )\right] \approx
B(B^o \rightarrow K^+ \pi^-)  - 2B(B^o\rightarrow K^o \pi^o)
\end{equation}
where  the result was expressed in terms of  branching ratios, denoted
by B() and $\tau^o/\tau^+$ denotes the ratio of the experimental $B^o$ and $B^+$ lifetimes.. This relation (\ref{eqapp}) states that the $I=3/2$ contributions to charged and neutral decays are equal. These $I=3/2$ contributions vanish in final $K\pi$ states produced by a penguin diagram which has isospin I=1/2.

The presence of an appreciable tree diagram contribution is indicated in new experimental data by the strong violation of the $I=1/2$ isospin relation between $B^+ $ and $B^o$ decays. The value of the following expression is predicted to vanish in a pure penguin transition differs from zero by many standard deviations.
\beq{newpuz2}
{{\tau^o}\over{\tau^+}}\cdot 2B(B^+ \rightarrow K^+ \pi^o) -
B(B^o \rightarrow K^+ \pi^- ) =
4.7 \pm 0.82 \not= 0
\eeq
where $B$ denotes the branching ratio in units of $10^-6$ and
we use the following experimental values

\beq{newtestex}
\begin{array}{ccl}
\displaystyle
B(B^o \rightarrow K^+ \pi^-)  =
19.4 \pm 0.6
 \hfill\\
\\
\displaystyle
{{\tau^o}\over{\tau^+}}\cdot B(B^+ \rightarrow K^o \pi^+) =
{{(23.1 \pm 1.0)}\over{1.07}}=
21.6\pm 0.93
\hfill\\
\\
\displaystyle
{{\tau^o}\over{\tau^+}}\cdot 2B(B^+ \rightarrow K^+ \pi^o)
= {{2\cdot (12.9\pm 0.6)}\over{1.07}}=
24.1\pm 0.56
\hfill\\
\\
\displaystyle
B(B^o\rightarrow K^o \pi^o) =
(9.4 \pm 0.6)
\end{array}
\end{equation}

On the other hand the I=3/2 components of individual  $B^+ $ and $B^o$ branching ratios are experimentally consistent with zero.
\beq{acpexp}
\begin{array}{ccl}
\displaystyle
B^o_{3/2} =  B(B^o \rightarrow K^+ \pi^-) - 2B(B^o\rightarrow K^o \pi^o) =
 (19.4 \pm 0.6)- 2\cdot (9.4 \pm 0.6) = 0.6 \pm 1.3
\hfill\\
\displaystyle
B^+_{3/2} =2B(B^+ \rightarrow K^+ \pi^o) -
B(B^+ \rightarrow K^o \pi^+ ) = (25.8\pm 1.2) - (23.1 \pm 1.0)
=2.7 \pm 1.6
\end{array}
\eeq
The suppression of the $I=3/2$ component is seen more clearly in the ratio of only 10\% between the $I=3/2$ contribution to neutral decays and the  $I=3/2$ contribution in the relation (\ref{newpuz2}) including both charged and neutral decays.
\beq{newtestx}
 \frac{2B(B^o\rightarrow K^o \pi^o) - B(B^o \rightarrow K^+ \pi^-)} {{{[\tau^o}/{\tau^+]}}\cdot[ B(B^+ \rightarrow K^o \pi^+) + 2B(B^+ \rightarrow K^+ \pi^o)] -
2B(B^o \rightarrow K^+ \pi^- )}
=0.09\pm 0.1
\end{equation}

This contrast between data on neutral and charged B decays
is surprising because the two decays differ only in the flavor of the spectator quark which does not participate in the weak interaction vertex. Furthermore, the strongest evidence against the I=3/2 contribution comes from the neutral decays where the presence of CP violation indicates that there must be a tree contribution and tree-penguin interference. This contrast is not predicted in any of the standard treatments \cite{nurosgro,ROSGRO,approxlip,approxgr,Gronau,ketaprimfix}.

\subsection {A search for spectator quark flavor dependence}

We first note that symmetries respected by QCD for the tree diagram can solve all puzzles.

The tree diagrams for the decays of a $\bar b$ antiquark and  a positive $B$ meson to a charmless final state are described by the
vertices
\beq{Bvert}
\bar b \rightarrow \bar q u \bar u; ~ ~ ~
B^+ =\bar b u \rightarrow \bar q u \bar u u
\end{equation}
where $\bar q$ denotes a  $\bar d$ antiquark for $\pi\pi$ decays or a
$\bar  s$ antiquark for $K\pi$ decays.

QCD describes the transition from this $\bar q u \bar u u$ initial state to
a final state of two color-singlet spinless mesons in an $S$ wave. We show below that all the puzzles are solved by an analysis using only the conservation laws respected by QCD without any detailed dynamics.

We first investigate the QCD symmetry constraints in a microscopic quark picture.

The tree diagram for  $B^{\pm}$ decay amplitudes contains two identical $u$ quarks in the final state which must obey the Pauli principle. The final state in $B^o$ decay
has no identical quarks and no Pauli constraints. This is the source of the dependence on the flavor of the spectator quark. For a rough approximation of the Pauli antisymmetrization needed in the final states  containing two $u$  quarks we simply discard
all amplitudes to final states containing two $u$ quarks.

Below we will show how using this approximation resolves the
puzzles and predicts new isospin relations that are in surprising agreement with experiment.

\begin{enumerate}
\item  The tree amplitude is predicted to  vanish in  $B^{\pm}$ decays
\begin{itemize}
\item Explains dependence on charge of spectator quark
\item Explains absence of CP violation in $B^{\pm}$ decays
\end{itemize}
\item  New isospin relations predicted between the $B\rightarrow K\pi$ amplitudes
\begin{itemize}
\item Individual $B^o$ and $B^+$ decay amplitudes satisfy pure $I=1/2$ penguin prediction(\ref{acpexp})
\item Strong violation of the $I=1/2$ only in relation between $B^+$ and $B^o$ decays(\ref{newpuz2}).
\end{itemize}
\item New experimental results show agreement with Pauli approximation predictions.
\begin{itemize}
\item Even in presence of tree contribution I=3/2 component experimentally consistent with zero
\item Penguin prediction of no I=3/2 fails in relations between $B^+$ and $B^o$ decays.
\end{itemize}
\end{enumerate}
\subsection {Theoretical loopholes}
But the theoretical basis of this Pauli approximation is questionable. The two $u$
quarks are separated in the final state and have different spatial wave functions. Furthermore this approximation fails experimentally when applied to $B\rightarrow \pi\pi$ amplitudes shown in fig. 2
 after changing the strange $\bar s$ antiquark to a
$\bar d$ antiquark. Here the final state also has two identical $u$ quarks but the tree amplitude does not vanish.

In the remainder of this paper we present a complete analysis of Pauli effects including symmetry constraints. We include entanglement that preserves these constraints even after the final state separates into two mesons. This treatment confirms the Pauli
suppression only for the tree diagram for  $B^{\pm}\rightarrow K\pi$ decays and not for the tree diagram in $B^{\pm}\rightarrow \pi\pi$ decays.

 \subsection {An independent derivation of  tree diagram suppression using only symmetries}

Without using the Pauli principle nor any microscopic quark model for hadrons we derive tree diagram suppression by applying only flavor symmetry and generalized Bose statistics to $B\rightarrow \pi^+ \pi^o$ decay. The final state of two isovector spinless bosons in the same isospin multiplet and in an S-wave must be symmetric in isospin. It must  be a pure isospin 2 state in a 27-dimensional representation of flavor SU(3). The I=1 combination is forbidden. These flavor-symmetry constraints remain even after the final state dissociates into two distant pseudoscalar mesons.

The $B\rightarrow \pi^+ \pi^o$ decay is described by a pure tree diagram with no penguin contribution.
The $B\rightarrow K^+ \pi$ decay also has a penguin diagram which is unrelated to  $B\rightarrow \pi^+ \pi^o$ decays. We limit our analysis here to the tree diagram contribution which is sufficient for our purposes. We now examine the $SU(3)_{flavor}$ ($d\rightarrow s$) transformation which changes $\pi^+(u\bar d)$ to $K^+(u\bar s)$ and changes the tree diagram $B^+ \rightarrow u\bar u \bar d u$ for $\pi \pi$ decays into the tree diagram $B^+ \rightarrow u\bar u \bar s u$ for $K \pi$ decays. It changes isospin to $V$ spin, the isospin triplet pion to a $V$ spin triplet and $\pi^+ \pi^o$ to $K^+V_{10}$
where $V_{10}$ denotes the neutral member of the $V$ spin triplet
\beq{V10}
V_{10} =  \frac {\pi^o +\sqrt 3\eta_8 }{2 }
\end{equation}
where $\eta_8$ is the SU(3) octet linear combination of $\eta$ and $\eta'$.

The quark compositions here are
\beq{pizero}
\pi^o = \frac {u \bar u - d\bar d}{\sqrt 2}; ~ ~ ~ V_{10} = \frac {u \bar u - s\bar s}{\sqrt 2} =  \frac {u \bar u -d \bar d +u \bar u +d \bar d -2s\bar s}{2 \sqrt 2}; ~ ~ ~
\eta_8 =
\frac {
u \bar u +d \bar d -2s\bar s}{\sqrt 6}
\end{equation}

We do not use this quark structure to derive tree diagram suppression, only symmetries.

The tree contribution to
$B\rightarrow K^+ \pi$ decay must have V=spin V=2. V=1 is forbidden.  The $K^+ \pi^o$ state can be seen to be only (1/4) of the SU(3) 27 related to the %
$B\rightarrow \pi\pi$ decay and only (1/4) of the V=2 state required by symmetry constraints. In the symmetry limit the strange analog of the $\pi^o\pi^+$ %
state in a pure SU(3) 27 is a $K^+ V_{10}$ state where $V_{10}$ denotes the $V$ spin analog of the $\pi^o$
with $V=1,V_z=0 $. The I=1 pion is mixture of V-spins 0 and 1 with probability of 1/4
for V=1. Only the V=1 component can combine with V=1 $K^+$ to produce V=2.

Thus  $SU(3)_{flavor}$ couplings transform the $\pi^+ \pi^o$  to a strange state in the 27-dimensional representation of flavor SU(3) which is mainly  $K^+\eta$ and $K^+\eta'$ with only a probability of (1/4) of being  $K\pi^o$.  The tree diagram contribution to charged B decays thus is reduced by a factor 4 which renders it negligibly small for our purposes here.
The remaining (3/4) of the $K^+ \pi^o$ state is classified in other representations of SU(3) which are not related to the $B\rightarrow \pi\pi$ decay.
Thus there is no possibility for using SU(3) with parameters determined  by $B\rightarrow \pi\pi$ decay to obtain parameters for analysis of $B\rightarrow K\pi$.
\section {Analysis with Pauli entanglement and color-spin}
\subsection{An approximate quantitative treatment of Pauli effects in $B \rightarrow K \pi$ decays.}

The tree diagram for the transition from an initial $B$ meson state  consisting of a $\bar b$ antiquark and a nonstrange spectator quark
to a strange charmless two-meson final state is written
\beq{secquanx}
\begin{array}{ccl}
\displaystyle
\ket{B_d}=\bar b d \rightarrow
\bar s  \cdot \left[u\bar u \right] d
\hfill\\
\ket{B_u}=\bar b u \rightarrow
\bar s  \cdot \left[ \kappa\cdot  u\bar u \right] u
\end{array}
\end{equation}
where the parameter $\kappa$ is a Pauli factor expressing the probability that the two u quarks are not Pauli blocked because they are not  in the same color-spin state.

We first consider the approximation $\kappa \approx 0$ where a $u$ quark produced by a weak interaction cannot enter the same state as a
$u$ spectator quark.
The states with $\kappa=0$ have no quark pairs of the same flavor. We call these Pauli-favored states.

When $\kappa=0$ the tree diagram (\ref{secquanx})is finite for neutral decays but vanishes in charged decays. This suppresses the tree contribution and CP
violation in  charged B decay while allowing the tree contribution in neutral decays.
This suppression is lost in conventional treatments which \cite{approxlip,approxgr,Gronau,ketaprimfix} separate the contribution of the tree diagram shown in fig. 2 into independent color favored and color suppressed diagrams shown in figs. 3 and 4. This separation introduces errors in the analysis of charged B decays by overlooking the symmetry constraints from the Pauli antisymmetry required by the presence of two identical u-quarks in the two diagrams. The two amplitudes go into one another under the exchange of the two u-quarks. The isospins of the u-quark produced in the weak vertex and of the spectator u quark must be entangled to always couple to isospin 1 even when the quarks are separated into two separate mesons.
We avoid these difficulties here by including entanglement in a general symmetry analysis

The tree amplitude produces a $u \bar u$ pair in the $b$ decay vertex. This amplitude is Pauli suppressed in charged $B$ decays where the spectator quark is also a $u$ quark. Thus in the $\kappa = 0$ approximation tree-penguin interference which might produce CP violation is present in neutral decays and absent in charged decays. This can explain how CP violation can be drastically changed by changing the spectator quark
and the otherwise mysterious result  (\ref{acp+}).

 In the remainder of this paper we show how the Pauli condition $\kappa \approx 0$ which seems surprising in the standard treatment\cite{approxlip,approxgr,Gronau,ketaprimfix} arises naturally from a detailed analysis including symmetry, the Pauli principle and entanglement. We now go beyond the $\kappa =0$ approximation and consider a full color-spin analysis.
\subsection {Detailed symmetry and Pauli analysis}
 All puzzles can  hopefully be resolved by a complete QCD calculation which is not yet feasible. However, we show below that symmetry methods are sufficient for obaining all relevant results without such QCD calculations.

In B decays to two pseudoscalar mesons a spin-zero state decays into two spin-zero particles with even parity and zero
internal orbital angular momentum. To conserve angular momentum the final state must
have no orbital angular momentum. We analyze the color, flavor and spin couplings in a tree diagram by separately determining
these couplings for the quark and antiquark pairs. We then combine these pairs to make a spin-zero color-singlet four-quark
state.

\begin {enumerate}
\item The state of the two u quarks and state of the two antiquarks $\bar u \bar q$ must both have wave functions
that are symmetric in space and flavor and antisymmetric in color-spin.
\begin{itemize}
\item The state of the two antiquarks $\bar u \bar q$ must have even parity since the odd parity state vanishes at the point where they are created.
\item Since parity is conserved only a flavor-symmetric component of the $uu$ state can combine with the even parity antiquark pair to produce an even parity final state.
\item Two u quarks in a spatially symmetric state are required by the Pauli
principle to be antisymmetric in color or spin.
\item The  antiquark  pair in (\ref{Bvert}) must also be antisymmetric in either color or spin.
\item Although no Pauli principle forbids a symmetric color - spin state for the antiquark pair
such states cannot combine with the $uu$ pair to make the spin-zero color singlet
final state $\pi\pi$ or $K\pi$.
\end{itemize}
\item The
fragmentation of a $uu\bar u \bar q$ state into a
$\pi^+\pi^o$ or $K^+\pi^o$ is a strong interaction which conserves flavor SU(3)
\begin{itemize}
\item The flavor SU(3) correlations remain entangled even after the final state dissociates into two separated mesons.
    \item The generalized Pauli principle requires both the $uu$ diquark and the $\bar u \bar q$ antidiquark
to be symmetric in flavor SU(3) and its  SU(2) subgroup isospin for $\pi\pi$ decays or
V-spin for $K\pi$ decays.
\item Each is therefore respectively in
the symmetric isospin state with $I=1$ or in the symmetric V-spin state with $V=1$
\end{itemize}
\item The
fragmentation of a $uu\bar u \bar q$ state into a
$\pi^+\pi^o$ or $K^+\pi^o$ is a strong interaction which conserves charge conjugation.
\item To combine charge conjugation and SU(3) Dothan\cite{dothan}  generalized the idea of G parity from SU(2) to SU(n).
\item For V spin Dothan parity $G_V$
defines the relative phases of the charge conjugate states in the same V-spin mu;tiplet
and defines the eigenvalue under charge conjugation of its C-eigenstate members.
\item We apply conservation of Dothan parity to G-parity and $G_V$ parity in the final state
\begin{itemize}
\item The pion isotriplet has isospin one and odd G parity.
The isoscalar pseudoscalar mesons $\eta$ and $\eta'$ have isospin zero and even G parity.
\item A nonstrange final state must be even under  charge conjugation  and have even
G parity to decay into two pions 
in an orbital S wave.
\item The $K^+$ and the three members of the V spin triplet have V spin 1 and odd $G_V$ parity.
\item The V-spin scalar and  vector pseudoscalar mesons are linear combinations of $\pi^o$, $\eta$ and $\eta'$
with V-spin zero, even $G_V$ parity and combinations with V-spin one, odd $G_V$ parity.
\item The $\pi^o$ is (3/4)
V-spin zero, even $G_V$ parity and (1/4) with V-spin one, odd $G_V$ parity.
\end{itemize}
\end {enumerate}

To produce a final state with even $G$ parity
the $(I=1,I_z=+1)$ diquark and the $(I=1,I_z=0)$ antidiquark must be coupled
symmetrically to  $(I=2,I_z=+1)$.
Similarly the $(V=1,V_z=+1)$ diquark and the $(V=1,V_z=0)$ antidiquark must be coupled
symmetrically to  $(V=2,V_z=+1)$ to produce a final state with even $G_V$ parity
.
These states are in the 27-dimensional representation of flavor SU(3).

The symmetry quantum numbers of the state at short times are assumed to be  preserved at longer times with entanglement if necessary. There is no need to consider the  difference between the spatial wave functions of the spectator b quark and the recoiling light meson. Their symmetries are entangled
even at large distances.

The dependence on spectator flavor arises from the Pauli blocking
by the spectator quark of a quark of the same flavor participating in the weak vertex. The u-quark produced by a
tree diagram is Pauli blocked by the spectator $u$ quark in $B^+$ decay but
is not affected by the spectator $d$ quark in neutral decays. This difference in Pauli
blocking suppresses the tree contribution and CP violation in charged $B$ decays but allows
tree-penguin interference and enables CP violation to be observed in neutral decays.

The final states in the 27 are produced from a $u$ quark pair in a color-spin state which
satisfies the Pauli Principle. Final states of two pseudoscalar mesons in other representations
of $SU(3)_{flavor}$ are Pauli suppressed.

A final $\pi^o\pi^+$ state is a pure $I=2$ state in a pure SU(3) 27.
Thus the tree diagram for the nonstrange transition
$(B^+ \rightarrow \pi+ \pi^o)$ is not Pauli suppressed.

A final $K^o\pi^+$ state
has no $V=2$ component, since both the $K^o$ and $\pi^+$
have V=1/2. Thus the tree diagram for the $K^o\pi^+$ decay must vanish and this
decay is pure penguin.

The final $K^+\pi^o$ state contains a $\pi^o$ which is a linear combination of
$V=0$ and $V=1$ states with probability of 1/4 for $V=1$.
The component with $V=0$ cannot combine with a $V=1$ $K^+$ to make $V=2$.
The $V=1$ component can   combine with a $V=1$ to make $V=2$
Thus the probability that the final
$K^+\pi^o$ state has a $V=2$ component is 1/4.
Thus we see that Pauli blocking suppresses the tree diagram for the
$(B^+ \rightarrow K^+ \pi^o)$
transition by a factor 4.

Present data are consistent with complete suppression but evidence for a
partial suppression is still down in the noise.

The $ud\bar u \bar s$ state created in the tree diagram for
$B_d$ decay has no such restrictions. It can be in a flavor SU(3)
octet as well as a 27. Its ``diquark-antidiquark" configuration includes the
flavor-SU(3) octet constructed from the spin-zero color-antitriplet
flavor-antitriplet ``good" diquark found in the  $\Lambda$ baryon and its
conjugate ``good" antidiquark. These ``good diquarks" do not exist in the
corresponding $uu\bar u \bar s$ configuration.

We again see that the Pauli effects produce a drastic dependence on spectator
quark flavor in the tree diagrams for $B \rightarrow K \pi$ decays.
Tree-penguin interference can explain both the presence
of CP violation in neutral decays and its absence charged decays.

We now note that the $ud$ pair in the final states must be isoscalar by the generalized Pauli principle. The final states must then be pure isospin eigenstates with $I=1/2$ and confirm the experimental result (\ref{acpexp}). In the standard treatments\cite{nurosgro,ROSGRO} the $I=3/2$ component is not suppressed in pure tree transitions

The quark diagrams for $B\rightarrow \pi\pi$ show $d \bar d $ states which are identified with a $\pi^o$.
However this turns the  $d \bar d$ state into an  $s \bar s$ state which is a linear combination of
$\eta$ and $\eta'$. There is no simple way to interpret the contribution of this diagram in a $K\pi$ final state.
We therefore have avoided the use of the simple tree diagrams in converting $B\rightarrow \pi\pi$ tree amplitudes to
$B\rightarrow K\pi$ tree amplitudes.

\section {Updating  conventional analysis using symmetry constraints and new data}

\subsection {Effects of Pauli constraint on tree diagram}
We now examine the effect of introducing symmetry constraints on a  detailed conventional analysis of new experimental data with no new theory.
Standard treatments \cite{nurosgro,ROSGRO,approxlip,approxgr,Gronau,ketaprimfix} of charmless B decays have incorrectly overlooked symmetry constraints in using data from
$B\rightarrow \pi\pi$ decays together with SU(3) flavor symmetry to obtain parameters
for analysis of $B\rightarrow K\pi$. At that time precise $B\rightarrow K\pi$ data were not yet
available. New more precise data revealed contradictions with this approach\cite{nuhuor1}.
The source of these contradictions can be seen in improper use of SU(3) and neglect of entanglement and the
Pauli principle. One problem arises because the $SU(3)_{flavor}$ transformation
which changes $\pi^+$ to $K^+$ changes the  $\pi^o$ into a linear combination of  $\pi^o$, $\eta$ and $\eta'$ and does not directly relate $B^{\pm} \rightarrow K\pi$ and  $B^{\pm} \rightarrow \pi \pi$.

Conventional analysis splits the amplitude for the tree diagram  shown in Fig. 2 into two components.
\begin {enumerate}
\item The color-favored tree amplitude, denoted by $T$ and shown in Fig. 3
\item The color-suppressed tree amplitude, denoted by $S$ and shown in Fig. 4
\end {enumerate}

The two amplitudes $T$ and $S$ are considered independent and the necessity for Pauli antisymmetry of states containing two identical u quarks is overlooked. This antisymmetry provides an additional symmetry constraint on these tree amplitudes which produces an approximate cancelation of the tree contribution. This cancelation explains the dependence of the tree diagram on the flavor of the spectator quark observed experimentally  in:
\begin {enumerate}
\item The difference in CP violation between neutral B decays (\ref{acp0}) and charged (\ref{acp+}) decays
\item The peculiar difference between isospins of tree diagram contributions shown in
(\ref{newpuz2}) and
(\ref{acpexp}).
\end {enumerate}

\subsection {Conventional analysis of difference rule experimental data support symmetry constraints}

Four experimental branching ratios
for  $B \rightarrow K\pi$ are available\cite{PDG,HFAG}.
The four final state
amplitudes are expressed in terms of three amplitudes\cite{approxlip,approxgr,Gronau,ketaprimfix}.
The two amplitudes $T$ and $S$ and the dominant gluonic penguin amplitude denoted by $P$ and shown in Fig. 1.

The standard treatment\cite{approxlip,approxgr,Gronau,ketaprimfix} neglects entanglement and assumes that  the two tree contributions are independent and are sufficiently small to enable the interference terms to taken only to first order,

\beq{pst}
\begin{array}{ccl}
\displaystyle
A[K^o\pi^+]=P; ~ ~ ~ A[K^+\pi^-]= T + P
\hfill\\
\\
\displaystyle
A[K^o\pi^o]={{1}\over{\sqrt{2}}} [S -  P]; ~ ~ ~
A[K^+\pi^o]={{1}\over{\sqrt{2}}} [T + S + P] 
\end{array}
\end{equation}

The four $B\rightarrow K\pi$ branching ratios
are determined by three parameters, $P$  and two interference terms $P\cdot T$ and $P\cdot S$.   But the new experimental data (\ref{acpexp}) show that both sides of (\ref{eqapp}) vanish as required by a pure penguin diagram. The data also reveal elsewhere (\ref{newpuz2}) a disagreement with the pure penguin and provide clues to the tree-penguin interference contribution.

We investigate these contradictions by investigating three different independent
differences between these branching ratios which eliminate the
penguin contribution. We express
(\ref{acpexp}) and (\ref{newpuz2})and in terms of the amplitudes $T$, $P$ and $S$:
 \beq{acpexpc2}
 \begin{array}{ccl}
\displaystyle
B(B^o \rightarrow K^+ \pi^-) - 2B(B^o\rightarrow K^o \pi^o) =
2\vec { P} \cdot (\vec T+\vec S)
 = 0.6 \pm 1.3  \approx 0
\hfill\\
2B(B^+ \rightarrow K^+ \pi^o) - B(B^+ \rightarrow K^o \pi^+ ) =
2\vec { P} \cdot (\vec T+\vec S)
=2.7 \pm 1.6  \approx 0
\hfill\\
{{\tau^o}\over{\tau^+}}\cdot 2B(B^+ \rightarrow K^+ \pi^o) -
B(B^o \rightarrow K^+ \pi^- ) =
 2\vec { P} \cdot (\vec S)
=4.7 \pm 0.82 \not= 0
 \end{array}
\eeq

Expressing the relation (\ref{newtestx}) in terms of the amplitudes $T$, $P$ and $S$ gives
\beq{newtest0}
\frac{\vec P\cdot (\vec T + \vec S)}{\vec P\cdot (\vec T - \vec S)}
=\frac{2B(B^o\rightarrow K^o \pi^o) - B(B^o \rightarrow K^+ \pi^-)} {{{[\tau^o}/{\tau^+]}}\cdot[ B(B^+ \rightarrow K^o \pi^+) + 2B(B^+ \rightarrow K^+ \pi^o)] -
2B(B^o \rightarrow K^+ \pi^- )}
=0.09\pm 0.1
\end{equation}

The new precise data show  that
the interference term between the dominant penguin amplitude
and the color-suppressed tree amplitude $\vec P \cdot \vec S $ is definitely
finite and well above the experimental errors. The sum
rule is still satisfied within two standard deviations and is now nontrivial.
But the interference term
$\vec P \cdot (\vec T+\vec S)$ is now equal to zero
well within the experimental errors (\ref{newtest0}).

There is no new theory here.
Choosing three independent differences in a way to minimize experimental errors
shows significant signals  well
above the noise of experimental errors that still fit an
overdetermination of the two parameters and lead to the result (\ref{newtest0}). The relation
between charged and neutral decays shows finite
tree-penguin interference contributions that can produce the observed
direct CP violation in neutral B-decays.  However the individual differences between two
neutral decays and between two charged decays satisfy the 1/2
pure penguin prediction of zero with no evidence for an I=3/2 contribution well below the noise and below the other
contribution. The absence of tree-penguin contributions in these differences is
completely unpredicted in the standard treatments when finite tree-penguin interference is detected in
both in the relation between charged and neutral decays.

This difference from the earlier conclusions \cite{approxlip,approxgr,Gronau,ketaprimfix}.
can be seen to arise from the Pauli principle  neglected  in conventional treatments.
The amplitudes $T$ and $S$  go into one another under
the interchange of the two identical $u$ quarks in $A[K^+\pi^o]$.

Consider the approximation $\kappa \approx 0$
where a $u$ quark produced by a weak interaction cannot enter the same state as a
$u$ spectator quark.
Pauli antisymmetry
requires $T$ and $S$ amplitudes to be equal and opposite.
We have shown that even in the general case, $\kappa \not =0$, this additional symmetry constraint reduces the tree contribution to $B^{\pm} \rightarrow K\pi$ by a factor 4 and makes it negligible here. Then $T + S \approx 0$ and we obtain

\beq{pstsq}
\begin{array}{ccl}
\displaystyle
|A[K^o\pi^+]|^2
=|\vec {{ P}}|^2; ~ ~ ~ |A[K^+\pi^-]|^2 \approx |\vec {{ P}}|^2+2\vec { P} \cdot \vec T
\hfill\\
\\
\displaystyle
2\cdot |A[K^o\pi^o]|^2
\approx |\vec {{ P}}|^2  + 2 \vec  P \cdot \vec T; ~ ~ ~
2\cdot |A[K^+\pi^o]|^2
\approx |\vec {{ P}}|^2
\end{array}
\end{equation}
where the approximate equalities hold to first order in the $T$
amplitude. These lead to the experimentally confirmed\cite{HFAG}
prediction (\ref{expred}) that the final states in both $B\rightarrow K\pi$ decays are
pure $I=1/2$ and that $I=3/2$ contributions
both vanish.
\beq{expred}
\begin{array}{ccl}
\displaystyle
|A[K^o\pi^+]|^2 \approx 2\cdot |A[K^+\pi^o]|^2
\hfill\\
\\
\displaystyle
A[K^+\pi^-]|^2 \approx 2\cdot |A[K^o\pi^o]|^2
\end{array}
\end{equation}
These predictions have been experimentally confirmed (\ref{acpexp}).

Thus tree-penguin interference with normally ignored Pauli effects
can explain the observed CP violation in
charged B-decays and its absence in neutral decays.
This shows how a nontrivial change in the weak
decay amplitude can arise from a change of the flavor of the spectator quark.

The two transitions which have a d-quark spectator have different final states  $K^+ \pi^-$ and $K^o \pi^o$. They are described individually (\ref{acpexpc2}) by the interference between the dominant common penguin diagram shown in
fig. 1 and color-favored and color-suppressed tree amplitudes shown respectively in
figs. 3 and 4. If tree-penguin is the source of the observed CP violation and the color-favored and color suppressed tree diagrams are independent one would expect no relation like (\ref{acpexpc2}) between
two independent amplitudes. The two transitions which have a u-quark spectator both lead to linear combinations of two different final states $K^+ \pi^o$ and $K^o \pi^+$ with different relative phase.  The
observation  of cancelation between the contributions of the two tree diagrams is surprising
in standard treatments missing Pauli entanglement. They assumed that the these two tree contributions
were completely independent and not expected to cancel.


\subsection {Effects of SU(3) symmetry breaking}

We note that $SU(3)_{flavor}$ seems to be badly broken. But the success of the Gell-Mann
Okubo mass formula indicates that simple first-order symmetry breaking can lead to useful
results.
The pion and kaon masses which are degenerate in the SU(3) limit are 140MeV and 494 MeV.
This is a large symmetry breaking. However if  we consider the SU(2) subgroups of U-spin and
V-spin the triplets of both include two kaons and a third neutral member.
The mass of the neutral members of the U spin and V spin triplet is
\beq{neutuv}
M(U_{10}) = M(V_{10}) =\frac{3\cdot M(\eta) +M(\pi)}{4} = 446 MeV
\eeq
This suggests that using each of the subgroups of isospin, U spin and V spin separately can give
reasonably approximate results. The large $K-\pi$ mass difference plays only a small role in these
subgroups. The serious breaking by the $K-\pi$ mass difference becomes important only when the three
subgroups are combined.

\section{A flavor topology analysis which includes final state interactions}
The unique flavor topology of the charmless strange quasi-two-body weak B decays
enables the results (\ref{newtest0}) to be obtained in a more general analysis
of these decays
including almost all possible diagrams including final state interactions and complicated
multiparticle intermediate states.

Consider diagrams for a charmless $B(\bar b q_s)$ decay into one strange and
one nonstrange meson, where $q_s$ denotes either a $u$ or $d$. The allowed
final  states must have the quark constituents  $\bar s n \bar n q_s$ where $n$
denotes a $u$ or $d$ nonstrange quark. We consider the topologies of all
possible diagrams in which a $\bar b$ antiquark and a nonstrange quark enter a
black box from which two final $q \bar q$ pairs emerge. We follow the quark
lines of the four  final state particles through the diagram going backward and
forward in time until they reach either the initial state or a vertex where
they are created. There are only two possible quark-line topologies for these
diagrams:

\begin{enumerate}

\item We call a generalized penguin diagram, shown in Fig. 1 , the sum of all possible diagrams in which a $\bar q q$ pair appearing in
the final state  is created by a gluon somewhere in the diagram. The quark
lines for the  remaining pair must go back to the weak vertex or the initial
state. This diagram includes not only the normally called penguin diagram but
all other diagrams in which the final  pair is created by gluons somewhere in
the diagram. This includes for example all diagrams normally called ``tree
diagrams" in which an outgoing $u \bar u$ or $c \bar c$ pair  is annihilated into gluons in a
final state interaction and a new isoscalar  $\bar q q$ pair is created by the
gluons.
There are two topologies for penguin diagrams.
\begin{itemize}
\item. A normal penguin diagram has a the spectator quark line continuing unbroken
 from the initial state to the final state.
 This  penguin contribution is described by a single parameter,
denoted by $P$ which is independent of the spectator quark flavor and contributes
equally to the $\bar s u \bar u q_s$ and $\bar s d \bar d q_s$ states.
\item A diagram which we call here an ``anomalous penguin" has the spectator
``u" quark in a $B^+$ decay annihilated in a final state interaction
against the $\bar u$ antiquark produced in a tree diagram. This diagram also   contributes
equally to the $\bar s u \bar u q_s$ and $\bar s d \bar d q_s$ states. But this
diagram denoted by $P_u$ is present only in
charged decays.
\end{itemize}
\item We call the ``tree diagram" the sum of all possible diagrams in which all
of the four quark lines leading to the final state go back to a initial $\bar s
u \bar u$ state created by the weak decay of the $b$ quark and the $ q_s$
spectator whose line goes back to the initial state.
     There are two possible couplings of the pairs to create final two meson
states from this diagram
 \begin{itemize}

\item The $\bar s u$ pair is coupled to make a strange meson; the  $\bar u q_s$
pair is coupled to make a nonstrange meson as shown in Fig.2. This is
conventionally called the  color-favored coupling. The contribution
of this coupling is described by a  single parameter, denoted by $T$.

\item The $\bar s q_s$ pair is coupled to make a strange meson; the  $u \bar u$
is coupled to make a nonstrange meson as shown in Fig. 3.
This is conventionally called the  color-suppressed coupling. The contribution
of this coupling is described by a  single parameter, denoted by $S$.

 \end{itemize}
 \end{enumerate}

All the results (\ref{newtest0}) obtained with the conventional definitions of $P$, $T$ and $S$ are seen to hold here
with the new definitions of $P$, $T$ and $S$. They now include contributions from
all final state interactions which conserve isospin and do not change
quark flavor.
The one final
state interaction not included is the $P_u$ diagram occurring in $B^+$ decays.
The flavor topology of this diagram creates an additional $I=1/2$ state
which is neglected in the derivation of the results (\ref{newtest0}).
These results hold as long  as the contribution of this $P_u$ diagram by final state interactions to the observed final states is negligible.

The additional $I=1/2$ contribution does not affect the ``difference rule" (\ref{eqapp}) which considers only the $I=3/2$ contributions.

In neutral $B_d$ decays there is no $P_u$ diagram.
Thus the simple relations (\ref{newtest0}) between the
$P$, $T$ and $S$ amplitudes hold for neutral decays are
not changed by isospin conserving final state interactions.

Further analysis of the contribution of this additional $I=1/2$ contribution is needed to include its modification of tree-penguin interference in obtaining definite values for CP violation in charged $B$ decay.

The electromagnetic penguin diagram is also  included in this flavor-topology formulation. The photon coupling to a $q \bar q$ pair can can be written
\beq{photon}
\gamma \rightarrow 2 u \bar u - d \bar d - s\bar s = 3u \bar u - [u \bar u + d \bar d + s\bar s]
\eeq
This coupling is included in the flavor-topology formulation as a linear combination of a tree coupling and a penguin coupling and contributes to the results (\ref{newtest0}). However there is now no simple relation between the $P$, $T$ and $S$ amplitudes and CKM matrices.

\section {Comparison with other approaches}

Previous analyses \cite{nurosgro,ROSGRO} were performed at a time when experimental
values for $B\rightarrow K\pi$ branching ratios were not
sufficiently precise to enable a significant test of the sum rule
(\ref{eqapp}). Values of each of the three interference terms in (\ref{newtest0})
were statistically consistent with zero.
The full analysis required the use of data from $B\rightarrow \pi\pi$ decays and
the assumption of $SU(3)_{flavor}$ symmetry. Contributions of the electromagnetic
penguin diagram were included and the relevant CKM matrix elements were included. But
there was no inclusion of constraints from the Pauli principle nor contributions from
final state interactions.

The present analysis uses new experimental data which enable a statistically significant
evaluation of the interference terms (\ref{newtest0}) without additional information
from $B\rightarrow \pi\pi$ decays or the assumption of $SU(3)_{flavor}$ symmetry relating
 $B\rightarrow K \pi$ and  $B\rightarrow \pi\pi$ decays.
Constraints from the Pauli principle are included in a general calculation including
color and spin and entanglement. Symmetries of the original weak amplitude are preserved  with
entanglement in the final state of two separated  mesons. Contributions from all isospin invariant finite state
interactions are included as well as constraints from the Pauli principle. The
flavor topology definition of the interference parameters includes contributions from the electromagnetic penguin diagram since the  quark states in final state of a photon can be rewritten as the sum of an isoscalar and
a $u\bar u$ state. However the flavor topology parameters are no longer simply related to the
CKM matrix elements. Additional assumptions and information are necessary to determine the CKM matrix elements
and explain CP violation.

The main advantage of this approach is that it gives simple explanations for the  absence of CP violation (\ref{acp+}) in charged B decays, the observed absence of an $I=3/2$ component in the final state,
and the vanishing of the experimental value
 (\ref{newtest0})

This vanishing of tree-penguin interference $B^+$ decays is explained without explicit  dynamics by a symmetry analysis including the constraints of the Pauli principle and entanglement on states containing a pair of identical $u$ quarks.
\section {Conclusion}

Experiment has shown that the penguin-tree interference contribution in $B^+
\rightarrow K^+\pi^o$ decay is very small and may even vanish. The
corresponding interference contributions to neutral $B\rightarrow K\pi$ decays
have been shown experimentally to be finite. The relation (\ref{newpuz2}) shows  that the $B\rightarrow K\pi$ transition is not a pure penguin.  The relation (\ref{acpexp}) shows  that the $I=1/2$ prediction by a pure  penguin diagram holds for the individual charged and neutral decays and is violated only in the ratio of the branching ratios for charged and neutral  decays.

The significant difference between the experimental values of expressions (\ref{acpexp}) and (\ref{newpuz2})
is not expected in the conventional analyzes. The two relating branching ratios for individual charged and neutral decays still vanish here while one relating charged and neutral case is finite. This indicates a surprising cancelation and motivates a search for a theoretical explanation.

We explain here that in charged decays the
previously neglected Pauli antisymmetrization  produces a cancelation
between color-favored and color-suppressed tree diagrams which differ by the
exchange of identical $u$ quarks. This smallness of penguin-tree interference
explains in a symmetry analysis without
QCD dynamics why CP violation
has been observed in neutral $B \rightarrow K\pi$ decays and not in charged
decays. Pauli cancelation does not occur in
neutral decay diagrams which have no pair of identical quarks.

\section*{Acknowledgements}

This research was supported in part by the U.S. Department of Energy, Division
of High Energy Physics, Contract DE-AC02-06CH11357. It is a pleasure to thank
Ben Grinstein, Michael Gronau, Yuval Grossman, Marek Karliner, Zoltan Ligeti, Yosef Nir,
Jonathan Rosner, J.G. Smith, and  Frank Wuerthwein for discussions and
comments.

%
\catcode`\@=11 
\def\references{
\ifpreprintsty \vskip 10ex
%
\hbox to\hsize{\hss \large \refname \hss }\else
\vskip 24pt \hrule width\hsize \relax \vskip 1.6cm \fi \list
{\@biblabel {\arabic {enumiv}}}
{\labelwidth \WidestRefLabelThusFar \labelsep 4pt \leftmargin \labelwidth
\advance \leftmargin \labelsep \ifdim \baselinestretch pt>1 pt
\parsep 4pt\relax \else \parsep 0pt\relax \fi \itemsep \parsep \usecounter
{enumiv}\let \p@enumiv \@empty \def \theenumiv {\arabic {enumiv}}}
\let \newblock \relax \sloppy
 \clubpenalty 4000\widowpenalty 4000 \sfcode `\.=1000\relax \ifpreprintsty
\else \small \fi}
\catcode`\@=12 



{\begin{figure}[htb]
$$\beginpicture
\setcoordinatesystem units <\tdim,\tdim>
\stpltsmbl
\putrule from -25 -30 to 50 -30
\putrule from -25 -30 to -25 30
\putrule from -25 30 to 50 30
\putrule from 50 -30 to 50 30
\plot -25 -20 -50 -20 /
\plot -25 20 -50 20 /
\plot 50 20 120 40 /
\plot 50 -20 120 -40 /
\springru 50 0 *3 /
\plot 120 20 90 0 120 -20 /
\put {$\overline{b}$} [b] at -50 25
\put {${q_s}$} [t] at -50 -25
\put {$\overline{s}$};[$\overline{d}$] [l] at 125 40
\put {$\overline{n}$} [l] at 125 -20
\put {$n$} [l] at 125 20
\put {${q_s}$} [l] at 125 -40
\put {$\Biggr\}$ $K(\vec k)$} [l] at 135 30
\put {$\Biggr\}$  $\pi(-\vec k)$} [l] at 135 -30
\put {$G$} [t] at 70 -5
\setshadegrid span <1.5\unitlength>
\hshade -30 -25 50 30 -25 50 /
\linethickness=0pt
\putrule from 0 0 to 0 60
\endpicture$$
\caption{\label{fig-2}} \hfill ``Gluonic penguin'' ($P$) diagram.
$G$ denotes any number of
gluons. $n$ denotes $u$ or $d$ quark.
\hfill~ \end{figure}}
{\begin{figure}[htb]
$$\beginpicture
\setcoordinatesystem units <\tdim,\tdim>
\stpltsmbl
\putrule from -25 -30 to 50 -30
\putrule from -25 -30 to -25 30
\putrule from -25 30 to 50 30
\putrule from 50 -30 to 50 30
\plot -25 -20 -50 -20 /
\plot -25 20 -50 20 /
\plot 50 20 120 40 /
\plot 50 -20 120 -40 /
\photonru 50 0 *3 /
\plot 120 20 90 0 120 -20 /
\put {$\overline{b}$} [b] at -50 25
\put {${q_s}$} [t] at -50 -25
\put {$\overline{u}$} [l] at 125 40
\put {$\overline{q_T};$} [l] at 125 20
\put {$u$}  [l] at 125 -20
\put {${q_s}$} [l] at 125 -40
\put {$\Biggr\}$
$I=1, V = 1$}  [l] at 135 30
\put {$\Biggr\}$ 
$I=1, V = 1$} [l] at 135 -30
\put {$W$} [t] at 70 -5
\setshadegrid span <1.5\unitlength>
\hshade -30 -25 50 30 -25 50 /
\linethickness=0pt
\putrule from 0 0 to 0 60
\endpicture$$
\caption{\label{fig-5}} \hfill Tree diagram. $\bar{q_T}$ denotes $\bar d$ or $\bar s$ antiquark.
\hfill~ \end{figure}}
{\begin{figure}[htb]
$$\beginpicture
\setcoordinatesystem units <\tdim,\tdim>
\stpltsmbl
\putrule from -25 -30 to 50 -30
\putrule from -25 -30 to -25 30
\putrule from -25 30 to 50 30
\putrule from 50 -30 to 50 30
\plot -25 -20 -50 -20 /
\plot -25 20 -50 20 /
\plot 50 0 120 -20 /
\plot 50 -20 120 -40 /
\photonru 50 20 *3 /
\plot 120 40 90 20 120 20 /
\put {$\overline{b}$} [b] at -50 25
\put {$q_s$} [t] at -50 -25
\put {$\overline{s}$};{$\overline{d}$} [l] at 125 40
\put {$u$} [l] at 125 20
\put {$\overline{u}$} [l] at 125 -20
\put {$q_s$} [l] at 125 -40
\put {$\Biggr\}$  $K(\vec k)$} [l] at 135 30
\put {$\Biggr\}$  $\pi(-\vec k)$}  [l] at 135 -30
\put {$W$} [t] at 70 15
\setshadegrid span <1.5\unitlength>
\hshade -30 -25 50 30 -25 50 /
\linethickness=0pt
\putrule from 0 0 to 0 60
\endpicture$$
\caption{\label{fig-4}} \hfill Color favored tree ($T$) diagram for $K \pi$ decay .
 \hfill~ \end{figure}}

{\begin{figure}[htb]
$$\beginpicture
\setcoordinatesystem units <\tdim,\tdim>
\stpltsmbl
\putrule from -25 -30 to 50 -30
\putrule from -25 -30 to -25 30
\putrule from -25 30 to 50 30
\putrule from 50 -30 to 50 30
\plot -25 -20 -50 -20 /
\plot -25 20 -50 20 /
\plot 50 20 120 40 /
\plot 50 -20 120 -40 /
\photonru 50 0 *3 /
\plot 120 20 90 0 120 -20 /
\put {$\overline{b}$} [b] at -50 25
\put {${q_s}$} [t] at -50 -25
\put {$\overline{u}$} [l] at 125 40
\put {$u$} [l] at 125 20
\put {$\overline{s};$} [l] at 125 -20
\put {${q_s}$} [l] at 125 -40
\put {$\Biggr\}$ $\pi(\vec k); \eta(\vec k); \eta'(\vec k)$}  [l] at 135 30
\put {$\Biggr\}$ $K(-\vec k)$} [l] at 135 -30
\put {$W$} [t] at 70 -5
\setshadegrid span <1.5\unitlength>
\hshade -30 -25 50 30 -25 50 /
\linethickness=0pt
\putrule from 0 0 to 0 60
\endpicture$$
\caption{\label{fig-5}} \hfill Color suppressed tree ($S$) diagram for $K \pi$ decay .
\hfill~ \end{figure}}
 \end{document}